\def\rfr#1{eq. (\ref{#1})}

\def\leti{Lense--Thirring}
\def\Rfr#1{Eq. (\ref{#1})}

\def\bar{\begin{eqnarray}}
\def\ear{\end{eqnarray}}

\def\eqi{\begin{equation}}
\def\eqf{\end{equation}}
\def\eqia{\begin{eqnarray}}
\def\eqfa{\end{eqnarray}}

\def\ct#1{\cite{#1}}
\def\lb#1{\label{#1}}

\def\oc2{$\mathcal{O}(c^{-2})$}

\documentclass[11pt]{article}
\usepackage{amsmath,amsthm,amscd,amssymb}
\usepackage{latexsym}
\usepackage{graphicx,epsfig}

\begin{document}

\noindent{\bf \LARGE{Some comments about  a recent paper on the
measurement of the general relativistic Lense-Thirring effect in
the gravitational field of the Earth with the laser-ranged LAGEOS
and LAGEOS II satellites}}
\\
\\
\\
{Lorenzo Iorio}\\
{\it Dipartimento Interateneo di Fisica dell' Universit${\rm
\grave{a}}$ di Bari
\\Via Amendola 173, 70126\\Bari, Italy
\\e-mail: Lorenzo.Iorio@libero.it}

\begin{abstract}
In this brief note some comments about the results presented in a
recently published paper on the measurement of the general
relativistic Lense-Thirring in the gravitational field of the
Earth are presented. It turns out that, among other things, the
authors might have yielded an optimistic evaluation of the error
budget because of an underestimation of the impact of the secular
variations of the even zonal harmonics of the geopotential. More
tests with real data by varying the observational time spans and
the magnitudes of $\dot J_4$ and $\dot J_6$ in the orbital
processor's force models should have been performed in order to
correctly address this important issue. Preliminary analytical
evaluations point towards a 15-45$\%$ error at 1-3$\sigma$ level,
respectively.
\end{abstract}
\section{On the adopted observable}
In a recently published paper \ct{ciu04}, submitted 2 June 2004
and accepted 10 September 2004, a measurement of the general
relativistic Lense-Thirring effect in the gravitational field of
the Earth with the laser-ranged satellites LAGEOS and LAGEOS II is
reported. The claimed total accuracy is $5-10\%$.  The observable
used in the analysis is the following linear combination of the
residuals of the longitudes of the ascending nodes $\Omega$ of
LAGEOS and LAGEOS II
 \eqi\delta\dot\Omega^{\rm LAGEOS}+ c_1\delta\dot\Omega^{\rm
LAGEOS\ II }\sim\mu_{\rm LT}48.2,\lb{iorform}\eqf in which
$c_1=0.546$ and $\mu_{\rm LT}$ is a scaling parameter which is 1
in the Einstein's General Theory of Relativity and 0 in the
Newtonian mechanics. \Rfr{iorform} allows to cancel out all the
static and time-dependent contributions of the first even zonal
harmonic coefficient $J_2$ of the multipolar expansion of the
Earth gravitational potential which represents one of the major
sources of systematic error. The terrestrial gravity model adopted
in \ct{ciu04} is the recently released GRACE-only EIGEN-GRACE02S
model \ct{02S}.

The possibility of using only the nodes of the LAGEOS satellites
in view of the improvements in the Earth gravity field solutions
from the CHAMP and GRACE missions was first presented in
\ct{rie03}, although not in an analytic and explicit form.
\Rfr{iorform} was explicitly published for the first time in
\ct{iormor04}. Then, it has been discussed in a number of other
papers \ct{ior03, ior04, iordorn04}.

Instead,  \rfr{iorform} is presented in \ct{ciu04} as an own
result of the authors who, not only miss out to correctly cite the
appropriate works \ct{rie03, iormor04, ior03, ior04, iordorn04},
but improperly refer to \ct{ciu86} (reference 19, pag. 959 of
\ct{ciu04} ). In that paper, which is almost twenty years old,
there is no mention of \rfr{iorform}. It is devoted to the well
known LAGEOS III mission in which the launch of a LAGEOS-type
satellite in an orbit with supplementary inclination with respect
to LAGEOS is presented. The goal of that configuration was to
cancel out exactly the contributions of all the even zonal
harmonics of the geopotential by using the simple sum of the nodes
as observable.

\section{Some possible criticisms on the error budget}
In this Section we will show that the error analysis presented in
\ct{ciu04} could be considered, perhaps, too optimistic, mainly
with respect to the impact of the gravitational errors which
represent the major source of systematic bias when \rfr{iorform}
is adopted.

In Section Error assesment, pag. 960 of \ct{ciu04} the authors
correctly assess the systematic error due to the static part of
uncancelled even zonal harmonics $J_{\ell}^{(0)}$ of geopotential:
indeed, it is $3-4\%$, according to the EIGEN-GRACE02S model. This
result agrees with that obtained in \ct{IOR04}. It should be
pointed out that these evaluations are at $1-\sigma$ level; at,
say, $3-\sigma$ we would get $9-12\%$. The first number (3$\%$)
comes from a root-sum-square calculation while the second number
(4$\%$) is the upper bound obtained by simply summing up the
individual error terms. The calibrated standard deviations of the
even zonal coefficients of EIGEN-GRACE02S have been used. A
possible criticism is that only this Earth gravity model has been
used in the presented analysis. Using different GRACE-only Earth
gravity models, like, e.g. EIGEN-GRACE01S (available at
http://op.gfz-potsdam.de/grace/results) and GGM01S \ct{ggm01s},
and analyzing the scatter of the so-obtained results would have
yielded better results in term of confidence and reliability.
Moreover, also the risk of using that particular model which gives
just the expected result would have been avoided.

Other problems may arise  when the authors show their $a\ priori$
error analysis for the time-dependent gravitational perturbations
(solar and lunar Earth tides, secular trends in the even zonal
harmonics of the Earth's field and other periodic variations in
the Earth's harmonics). Indeed, they claim that, over an
observational time span of 11 years, their impact would be 2$\%$.
This evaluation is based on reference 30, pag. 960 of \ct{ciu04}
which refers to the WEBER-SAT/LARES INFN study; it has nothing to
do with the present node-only LAGEOS-LAGEOS II combination. On the
contrary, many recent studies \ct{ries03,2,3,4,5,6} mainly
focussed on the gravitational part of the error budget in the
performed or proposed
\leti\ tests with LAGEOS-like satellites are not even included
in the attached .doc file which should overcome the unavoidable
space limitations posed by the Letter format\footnote{On the
contrary, a large number of references are devoted to the
non-gravitational perturbations which, instead, play a minor role
in this case due to the small senistivity of the LAGEOS nodes to
them.}. Moreover, this estimate may turn out to be optimistic
because of the secular variations of the even zonal
harmonics\footnote{The problem of the secular variations of the
even zonal harmonics in post-Newtonian tests of gravity with
LAGEOS satellites has been quantitatively  addressed for the first
time in \ct{luc03}. In regard to the \leti\ measurement with
\rfr{iorform}, it has been, perhaps, misunderstood in
\ct{iormor04}. } $\dot J_{\ell}$. Indeed, \rfr{iorform} allows to
cancel out $\dot J_2$, but is sensitive to $\dot J_4$, $\dot
J_6$,..., as pointed out in \ct{IOR04}. The uncertainties in the
$\dot J_{\ell}$ are still quite large. On the other hand, their
impact on the \leti\ measurement grows linearly in time. Indeed,
the mismodelled shift, in mas, of \rfr{iorform} due to the secular
variations of the uncancelled even zonal harmonics can be written
as \eqi \sum_{\ell=2}\left(\dot\Omega_{.\ell}^{\rm LAGEOS}+c_1\
\dot\Omega_{.\ell}^{\rm LAGEOS\ II }\right)\frac{\delta\dot
J_{\ell}}{2}T^2_{\rm obs},\lb{quadr}\eqf where the coefficients
$\dot\Omega_{.\ell}$ are $\partial \dot\Omega_{\rm class}/\partial
J_{\ell}$ and have been explicitly calculated up to degree
$\ell=20$ in \ct{3}. It must be divided by the gravitomagnetic
shift, in mas, of \rfr{iorform} over the same observational time
span \eqi \left(\dot\Omega_{\rm LT}^{\rm LAGEOS}+0.546\
\dot\Omega_{\rm LT}^{\rm LAGEOS\ II} \right) T_{\rm obs}=48.2\
{\rm mas\ yr^{-1}}\ T_{\rm obs}.\eqf
 By assuming $\delta\dot J_4=0.6\times 10^{-11}$
yr$^{-1}$ and $\delta\dot J_6=0.5\times 10^{-11}$ yr$^{-1}$
\cite{cox02}, it turns out that the  percent error on the
combination \rfr{iorform} grows linearly with $T_{\rm obs}$ and
would amount to $1\%$ over one year at $1-\sigma$ level. This
means that, over 11 years, their impact might range from 11$\%$
(1-$\sigma$) to 33$\%$ (3-$\sigma$). Alternatively, if we look at
the rate\footnote{Indeed, the normalized slope of the time series
is measured. }, in mas yr$^{-1}$, these figures must be doubled.
Indeed, the mismodelled secular rate due to the $\dot J_{\ell}$ is
\eqi \sum_{\ell=2}\left(\dot\Omega_{.\ell}^{\rm LAGEOS}+c_1\
\dot\Omega_{.\ell}^{\rm LAGEOS\ II }\right)\delta\dot
J_{\ell}T_{\rm obs},\lb{dota}\eqf which must be divided by the
\leti\ secular trend \eqi \dot\Omega_{\rm LT}^{\rm LAGEOS}+0.546\
\dot\Omega_{\rm LT}^{\rm LAGEOS\ II} =48.2\ {\rm mas\
yr^{-1}}.\eqf This subtle and important point should have been
addressed with tests with real data by varying the magnitudes of
$\dot J_{2}$ and $\dot J_4$ in the force models of the orbital
processor over different observational time spans.

Another controversial point is that it is unlikely that the
various errors of gravitational origin can be summed in a
root-sum-square way because of the unavoidable correlations
between the various phenomena of gravitational origin. It would be
more conservative to add them. In this case, the
$(J_{\ell}^{(0)}-\dot J_{\ell})$ error would range from 15$\%$
(4$\%$+11$\%$) at $1-\sigma$ level to 45$\%$ (12$\%$+33$\%$) at
$3-\sigma$ level over a 11-years long observational time
span\footnote{The evaluations of \rfr{quadr} have been used; if
\rfr{dota} is used the upper bounds become $4\%+22\%=26\%$
(1-$\sigma$) and $12\%+66\%=78\%$ (3-$\sigma$).}. The so obtained
global gravitational error can be added in quadrature to the
non-gravitational error. Even by assuming the 2$\%$ authors'
estimate of the time-dependent part of the gravitational error,
the upper bound errors would be $\sqrt{(4+2)^2+2^2}\%=6\%$ at
1-$\sigma$, $\sqrt{(8+4)^2+4^2}\%=13\%$ at 2-$\sigma$ and
$\sqrt{(12+6)^2+6^2}\%=19\%$ at 3-$\sigma$. Instead, at the end of
the Section Total uncertainty, pag. 960 of \ct{ciu04} and in their
Supplementary Information .doc file the authors add in quadrature
the doubled error due to the static part of the geopotential (the
2$\times 4\%$ value obtained from the sum of the individual error
terms), their perhaps optimistic evaluation of the error due to
the time dependent part of the geopotential and the
non-gravitational error getting $\sqrt{8^2+4^2+4^2}\%=10\%$ at
2-$\sigma$. On the other hand, in the Supplementary Information
.doc file it seems that they triple the 3$\%$ error due to the
static part of the geopotential obtained with a root-sum-square
calculation  and add it in quadrature to the other (not tripled)
errors getting $\sqrt{9^2+2^2+2^2}\%\leq 10\%$ at 3-$\sigma$.
These calculations look like tricks to get just a desired value,
i.e. 10$\%$.

Finally, it is hard to understand why the authors very often refer
to the LAGEOS-LARES proposed experiment and to the related
simulations and error budgets. It is rather confusing and
misleading. The LAGEOS-LAGEOS II combination of \rfr{iorform} is,
by construction, designed in order to exactly cancel out the $J_2$
term with an approach which can be applied to any orbital
configuration given a pair of satellites in different orbits or a
pair of different Keplerian orbital elements of the same
satellite. On the contrary, the observable originally proposed for
the LAGEOS-LARES mission is the simple sum of their nodes. If the
orbital parameters of LARES, quoted in Table \ref{para}, were
exactly equal to their nominal values, all the even zonal
harmonics would be exactly cancelled out. Instead, the sum of the
nodes would be affected, to a certain extent, by the whole range
of the even zonal harmonics of the geopotential due to unavoidable
departures from the LARES nominal configuration because of the
orbital injection errors and mission design (the eccentricity of
LARES would be one order of magnitude larger than that of LAGEOS),
i.e. the coefficients of the classical nodal precessions would not
be exactly equal and opposite $\dot\Omega_{.\ell}^{\rm LAGEOS}\neq
-\dot\Omega_{.\ell}^{\rm LARES}$ for $\ell=2,4,6,8,...$. In the
Supplementary Information .doc file the combination of the nodes
of LAGEOS and LAGEOS II of \rfr{iorform} is presented as if it is
only slightly different with respect to the sum of the nodes of
the originally proposed LAGEOS-LARES configuration, apart from a
18 deg offset in the inclination of LAGEOS II with respect to
LARES. The differences in the eccentricities and the semimajor
axes, which do play a role \ct{2}, have been neglected.
{\small\begin{table}\caption{Orbital parameters of the existing
LAGEOS and LAGEOS II and of the proposed LARES.  }\label{para}

\begin{tabular}{llll}
\noalign{\hrule height 1.5pt}

Satellite & $a$ semimajor axis (km) & eccentricity $e$ & inclination $i$ (deg)  \\

\hline

LAGEOS    &  12270    & 0.0045 &  110\\
LAGEOS II &  12163    & 0.0135 & 52.64\\
LARES    & 12270    & 0.04  & 70.0 \\

\noalign{\hrule height 1.5pt}
\end{tabular}

\end{table}}

\section{Conclusions}
The main objections to the results presented in \ct{ciu04} can be
summarized as follows
\begin{itemize}
  \item The authors attribute to themselves the combination of \rfr{iorform} used in their analysis
  and consistently ignore almost all the works of
  other scientists on the gravitational part of the error budget
  which, in this case, represents the major source of systematic
  bias because of the low sensitivity of the nodes of the LAGEOS satellites to the
  non--gravitational perturbations.

  \item No different GRACE-only Earth gravity models have been
  used to support the error assessment.

  \item A major point is represented by the impact of the secular
  variations of the uncancelled even zonal harmonics $\dot J_4$,
  $\dot J_6$. Their mismodelling  might induce errors as large as tens percent over
  an observational time span of 11 years, according to
  preliminary analytic analyses.
  Tests with real data by varying the observational time spans and
  the magnitudes of $\dot J_4$,
  $\dot J_6$ in the orbital processor's force models should have been extensively conducted.
  The authors present a 2$\%$ estimate of the global
  time-dependent part of the gravitational error and support it
  with an improper reference to the preparatory study
  for the WEBER-SAT/LARES mission submitted to the Italian National Institute of Nuclear Physics
  (INFN).

  \item The static and time-dependent gravitational errors should
  be simply added together because of the correlations among the various
  gravitational phenomena. Instead, in \ct{ciu04} they are summed in quadrature
  and presented in a way which seems a ad hoc trick just to get a
  10$\%$. Even by assuming the authors' 2$\%$ estimate of the
  time-dependent part of the gravitational error, the 3-$\sigma$
  upper bound error would amount to 19$\%$, contrary to the $10\%$
  value presented in the Supplementary Information .doc file.
\end{itemize}

\section*{Acknowledgments}
I am grateful to J. Ries (Center for Space Research, CSR) for
useful comments and discussions about the use of different Earth
gravity models.


\end{document}